# Automatic detection of estuarine dolphin whistles in spectrogram images

O. M. Serra [a,*], F. P. R. Martins [b], L. R. Padovese [c]

[a] *Universidade de São Paulo, Escola Politécnica, Department of Mechanical Engineering, Av. Prof. Mello Moraes, 2231, zip 05508-030, São Paulo, Brazil*
[b] *Universidade de São Paulo, Escola Politécnica, Department of Mechanical Engineering, Av. Prof. Mello Moraes, 2231, zip 05508-030, São Paulo, Brazil*
[c] *Universidade de São Paulo, Escola Politécnica, Department of Mechanical Engineering, Av. Prof. Mello Moraes, 2231, zip 05508-030, São Paulo, Brazil*

**Abstract**

An algorithm for detecting tonal vocalizations from estuarine dolphin (*Sotalia guianensis*) specimens without interference of a human operator is developed. The raw audio data collected from a passive monitoring sensor in the Cananéia underwater soundscape is converted to spectrogram images, containing the desired acoustic event (whistle) as a linear pattern in the images. Detection is a four-step method: first, ridge maps are obtained from the spectrogram images; second, a probabilistic Hough transform algorithm is applied to detect roughly linear ridges, which are adjusted to the true corresponding shape of the whistles via an active contour algorithm; third, feature vectors are built from the geometry of each detected curve; and fourth, the detections are fed to a random forest classifier to parse out false positives. We develop a system capable of reliably classifying roughly 97% of the characteristic patterns detected as *Sotalia guianensis* whistles or random empty detections.

## 1. Introduction

The estuarine dolphin (*Sotalia guianensis*), also known as Guiana dolphin, is one of the smallest species of dolphin, and is native to the Atlantic coastal regions of Central and South America, ranging from Honduras to the state of Santa Catarina, Brazil (Flach et al., 2008).

It is a species that lives in cohesive groups, commonly ranging from 2 to 6 individuals, although groups of over 50 specimens have been observed (Monteiro Filho, 2008). Due to residing in coastal areas, the estuarine dolphin is particularly susceptible to human impacts, with the greatest threats to the species being pollution, habitat degradation, accidental capture and increase in the frequency of maritime traffic (ICMBio, 2011).

Information on the populational and social dynamics of the estuarine dolphin is currently scarce. The data entry for this species in the International Union for Conservation of Nature (IUCN) Red List was classified as "data deficient" (DD) until 2017 due to lack of conclusive data. It has since been classified as "near threatened" (NT), with the rating being changed in 2018 (Secchi et al., 2018).

One of the vocalizations emitted by the estuarine dolphin is a tonal sound (whistle) that may be useful in detecting and categorizing its specimens within a given region. Dolphin whistles and their shapes in spectrogram images were characterized in (Bazúa-Durán, 2004; Henderson et al., 2011; Kriesell et al., 2014). In these papers, the visual patterns corresponding to the whistles roughly resemble line segments, or combinations thereof.

As will be detailed in section 2, the data used in this research contains several estuarine dolphin vocalizations as object of study, and is part of a soundscape of maritime protected regions in southeastern Brazil done in (Sánchez-Gendriz and Padovese, 2016). Our current research is then concerned with the automatic detection of the aforementioned estuarine dolphin vocalizations from raw audio data. The source data was converted to spectrogram images, which were used thereafter to identify the vocalizations based on their corresponding patterns in the images.

Features from the desired sounds can be extracted by audio analysis (Malfante et al., 2018), knowledge of the geography of the data collection sites (Henderson et al., 2011) or the geometry of the spectrogram images (Esfahanian et al., 2013; MacBride and Roth, 2016).

An algorithm for detection of sounds from fish choruses based only on audio analysis was implemented in (Malfante et al., 2018), relying solely on auditory patterns. This method had very high accuracy for well-defined conditions, but encountered problems when the selected time window contained multiple detections from different classes.

Erbs et al. (2017) and Henderson et al. (2011) propose methods based on the statistical distribution of features from the desired acoustic events for classifying whistle sounds from different species, and present high accuracy rates for inter- and intra-species classification.

Image segmentation algorithms are widely employed in the literature (Gillespie, 2004; Kershenbaum and Roch, 2013; Mellinger et al., 2011). Further classification of the identified sounds is often done with random forest classifiers (Henderson et al., 2011) or methods based on eigenfunction problems (MacBride and Roth, 2016; Karnowski and Johnson, 2015).

* Corresponding author. Tel: +55-011-3091-9648; fax: +55-011-3091-5687;
E-mail: otaviomserra@usp.br

A similar algorithm to the one employed in our article is detailed in (Gillespie, 2004) for the detection of right whale calls. Edge maps were obtained from smoothened spectrogram images, from which outlines of the desired events were determined. These events were then run through a classifier to reliably detect whether they corresponded to the right whale calls or not.

Kershenbaum and Roch (2013) developed a method for the identification of cetacean tonal sounds via ridge maps on a spectrogram image, as is done in this study. Their method took the ridges detected in the spectrogram images as potential detections, and tallied the number of false positives in comparison to an earlier algorithm.

The method proposed in the current article is a mixed approach if compared to the most recent methods in the literature. In this article, we develop an algorithm to detect estuarine dolphin vocalizations through spectrogram image analysis. The computational vision method employs an image segmentation method based on the Hough transform (Duda and Hart, 1972) and active contours. Further classification is done via a random forest classifier.

This article has the following structure: section 2 details the process of data collection and the algorithms employed in this study; in section 3, we present the most important patterns observed and the final performance of the classification algorithm; in section 4, we present our conclusions over the study.

## 2. Materials and methods

The dataset used in this research was obtained following the article published in (Sánchez-Gendriz and Padovese, 2016). The acoustic signals were recorded at a depth of 8m underwater via an autonomous passive monitoring system (OceanPod), developed at LACMAM (Laboratory of Acoustic and Environment of the Escola Politécnica da Universidade de São Paulo) (see Fig. 2.1).

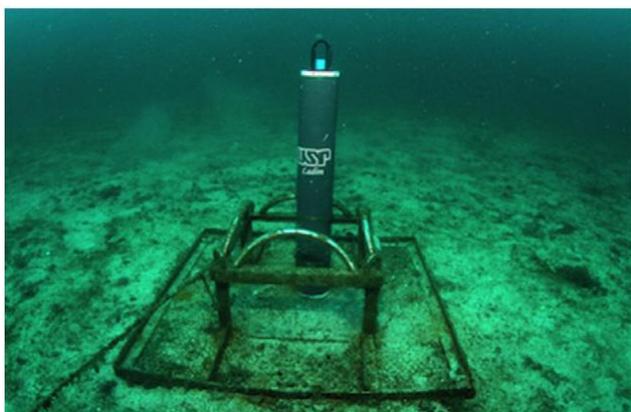

Fig. 2.1. OceanPod 1.0 used for recording. Source: Sánchez-Gendriz and Padovese (2016)

The data acquisition campaign was conducted at a bay close to Cananeia (Fig. 2.2), a city in the south of São Paulo State, Brazil from February 07 to 10, 2017. The recording was carried out at a sampling frequency $f_s = 48 kHz$, 16 bit resolution, and the grabbed data were continuously stored in an SD card, as a sequence of .wav files, each with size corresponding to 3 minutes of acoustic emission. The sensitivity of the system was set to −150 dB re 1V/μPa and its frequency band from 10 Hz to $f_s/2$. During the campaign, 1460 .wav files, totalizing 23,4 GB, were recorded; from these, 110 were randomly selected to feed the algorithms presented in this article.

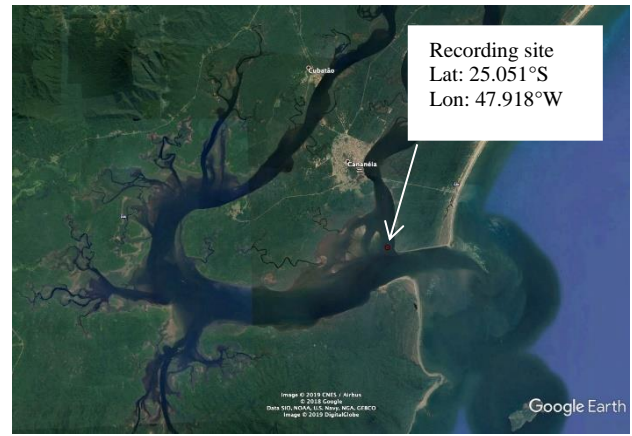

Fig 2.2: Location of the recording site. Source: (C) Google Earth. [color]

As will be detailed in the following subsections, the methodology proposed for detecting vocalizations of estuarine dolphin specimens was based on the application of image processing and machine learning techniques to the recorded signal spectrograms. Python 3.6 (Python Software Foundation) was used as a programming language. Additional Python libraries and resources employed were: Numpy (Oliphant, 2006), Matplotlib (Hunter, 2007), Pandas (McKinney, 2010), scikit-learn (Pedregosa et al., 2011), scikit-image (van der Walt et al., 2014), PyDub (https://pypi.org/project/pydub/) and Jupyter (Kluyver et al., 2016).

### 2.1. Overview of the algorithm

The structure of the algorithm for recognizing estuarine dolphin vocalizations subdivides naturally into four steps: signal and image pre-processing, image segmentation, computation of feature vectors and classification. A simplified framework of the algorithm is shown in Fig. 2.1.1.

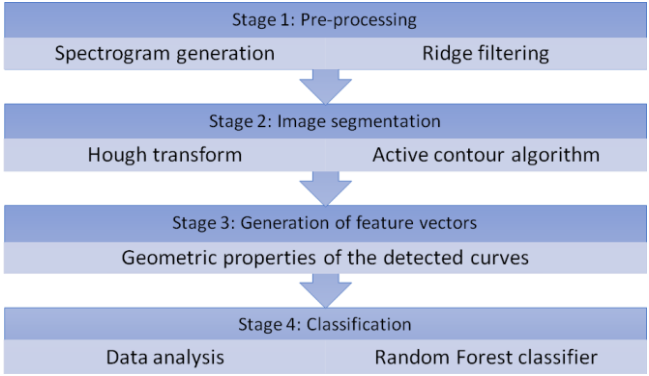

Fig. 2.1.1: Proposed algorithm for detecting estuarine dolphin vocalizations.

Each of the 110 audio files selected was partitioned into sixty 3-second snippets, each of them being turned into a spectrogram image following the Matplotlib implementation. The 3-second window was a duration deemed appropriate for the visualization of the desired pattern. The dolphin vocalizations of interest form curves that roughly resemble line segments, observed to make angles with the horizontal axis within the interval from 15° to 75°. Fig. 2.1.2 shows an example of this typical pattern.

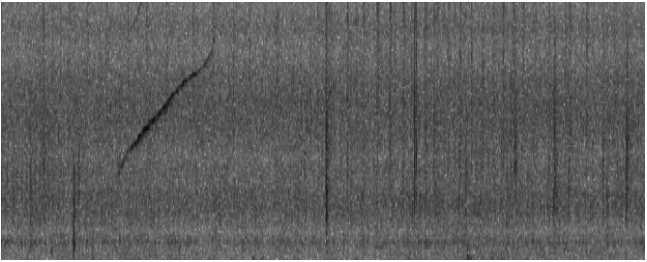

Fig. 2.1.2: Example of a roughly linear pattern corresponding to the acoustic event of interest.

A general summary of the algorithm is as follows. We detect possible instances of the desired pattern, via a Hough transform (Duda and Hart, 1971) on the intensity ridge map of the image. An active contour algorithm (Kass et al., 1988) is thereby employed to adjust the lines to the true shape of the pattern, obtaining open-ended curves. Geometric features are extracted from each curve and fed to a machine learning algorithm, which finally determines whether the curve corresponded to an event of interest.

**2.2. Image pre-processing**

The first step in detecting the characteristic patterns of interest in a spectrogram image is to make such features stand out. The chosen method was the Frangi vesselness filter (Frangi et al., 2006), a ridge detection algorithm that singles out regions of the image with large dark patches.

As detailed in (Fu et al., 2017), the Frangi filter excels in detecting tubular shapes in structures, and is mostly used in medical research. The unique, roughly linear structure of the estuarine dolphin whistles means that a filter capable of singling out tubular-shaped regions will produce good results in our spectrogram images.

If $H$ is the Hessian matrix of the spectrogram image, we compute its eigenvalues $\lambda_{1,2}$, with numbering conventioned as $|\lambda_2| > |\lambda_1|$. The vesselness value is then calculated from the eigenvalues, and depends on the standard deviation $\sigma$ used to compute the Hessian. Eq. 2.2.1 gives the output of the Frangi filter.

$$V_0(\sigma) = \begin{cases} 0, if\ \lambda_2 < 0; \\ e^{-\frac{R_B^2}{2\beta^2}}\left(1 - e^{-\frac{S^2}{2c^2}}\right), otherwise. \end{cases} \quad (2.2.1)$$

In Eq. 2.2.1, $S = \sqrt{\lambda_1^2 + \lambda_2^2}$, $R_B = \|\lambda_1\|/\|\lambda_2\|$ and $\beta, c$ are parameters. The scikit-image implementation used in this research follows the original Frangi filter.

Then, the resulting map was binarized through a manually chosen threshold for $V_0$, separating low- from high-intensity regions. An example of this process is shown in Fig. 2.2.1.

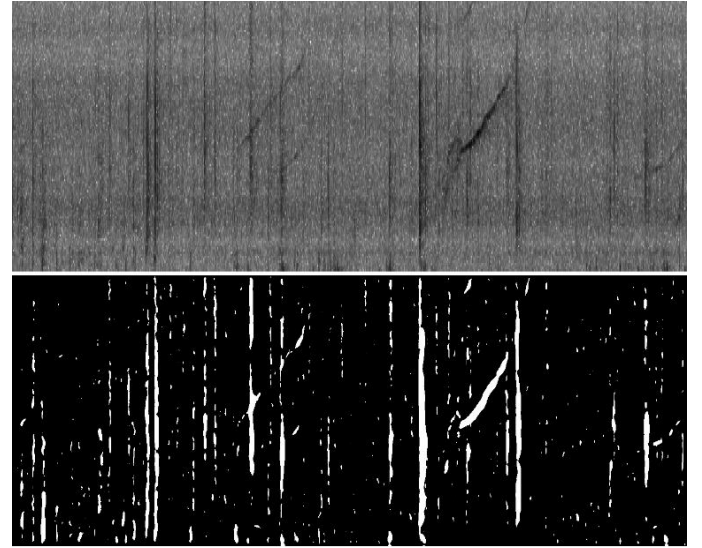

Fig. 2.2.1: (top) Original image; (bottom) Respective binarized Frangi map.

**2.3. The Hough transform**

In order to detect linear patterns in the binarized Frangi maps, we applied a Hough transform algorithm. As described in (Duda and Hart, 1972), a straight line $L$ (Fig.2.3.1) of image space xy, described in polar coordinates by

$$r = x\cos\theta + y\sin\theta \quad (2.3.1)$$

is transformed into a point $P$ with coordinates $(r, \theta)$ in Hough space.

Furthermore, as the number of image points incident on L corresponds to the "intensity" of the respective point P in the Hough map, these points of greatest intensity are the ones more likely to represent real straight lines in the image space.

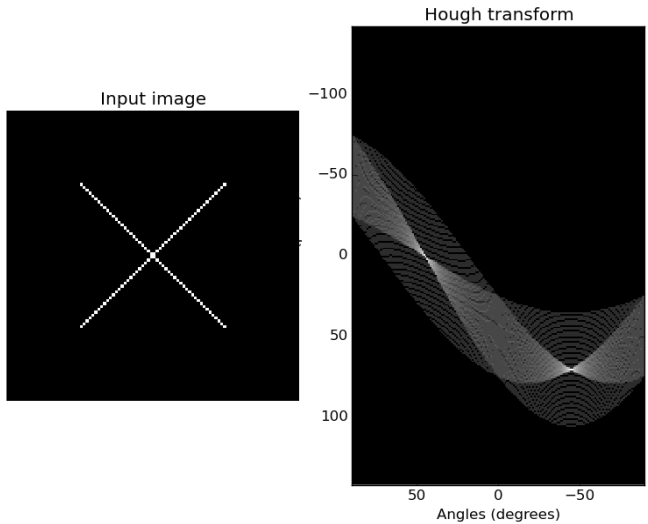

Fig. 2.3.1: The Hough parameter space for two straight lines. Source: scikit-image

The particular implementation employed in this research is from scikit-image, which is based on (Galambos et al., 1999). It applies a progressive probabilistic method to determine where a line starts and ends, in addition to verifying whether it exists or not. An example of this method, applied to the same source image as Fig. 2.2.1, is shown in Fig. 2.3.2.

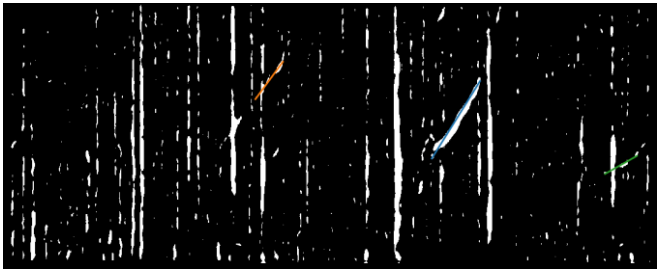

Fig. 2.3.2: Lines (in color) detected using the probabilistic Hough transform. [color]

## 2.4. The active contour (*snake*) algorithm

Since the locations of the vocalizations of interest in the spectrogram images are roughly located by the Hough Transform, an active contour algorithm is applied in order to accurately capture the shape of those patterns.

The resulting curves overlaying the referred objects will henceforth be referred to as "*snakes*," in agreement with the literature. A *snake* is defined to be a piecewise smooth spline
$$v(s) = (x(s), y(s)), s \in [0,1].$$
The scikit-image implementation is based on (Kass et al., 1988). The main points of this algorithm are presented in the following paragraphs.

The straight lines obtained previously, e.g. the ones in Fig. 2.3.2, are used as the first approximate solution for the variational problem implemented by the *snake* algorithm. The traditional algorithm searches iteratively for local minima of the energy functional associated to the *snake*, given in Eq. 2.4.1 in accordance with (Kass et al., 1988).

$$E_{snake}^* = \int_0^1 E_{int}(v(s)) + E_{image}(v(s)) + E_{con}(v(s)) \, ds \quad (2.4.1)$$

Interpretations for the terms are as follows. The internal energy $E_{int}(v(s))$ of the *snake* is defined in Eq. 2.4.2.

$$E_{int}(v(s)) = \frac{1}{2}\left(\alpha(s)\left|\frac{\partial v}{\partial s}(s)\right|^2 + \beta(s)\left|\frac{\partial^2 v}{\partial s^2}(s)\right|^2\right) \quad (2.4.2)$$

It imposes a piecewise smoothness constraint. The second term $E_{image}(v(s))$, commonly referred to as image energy, consists of conditions that attract the *snake* to desired patterns in the image, e.g. edges or ridges. A formulation of this term for edge attraction is given in Eq. 2.4.3, assuming that detection of terminations of edges is not important, due to endpoints remaining fixed.

$$E_{image}(v(s)) = w_1 I(v(s)) + w_2 |\nabla I(v(s))|^2 \quad (2.4.3)$$

In Eq. 2.4.3, $I(x, y)$ is the gray level image function at the point $(x, y)$. The constraint energy $E_{con}(v(s))$ keeps the *snake* close to a desired local minimum of its energy functional. The definition of this term is situational and, in our research, it can be considered negligible.

In short, the algorithm attracts the *snake* to a desired feature of the image, while still maintaining its piecewise smoothness. In computational applications, a *snake* is a discretization of the analytic approach showed above, taking as a first estimate the straight line obtained via the Hough transform, with N equally spaced points. This piecewise line segment then undergoes a numerical minimization of its energy functional, given by Eq. 2.4.1.

The scikit-image implementation takes the first- and second-order functional coefficients $\alpha(s)$ and $\beta(s)$ from Eq. 2.4.2 to be constants defined by the user. In some cases, it may be desirable to let one of the parameters equal zero at certain values of $s$. For instance, a value of $\beta = 0$ would allow for a second-order discontinuity (a sharp turn) at that point.

In our work, although $\alpha$ and $\beta$ were defined as constants, little emphasis was given to second-order continuity, resulting in a low value for $\beta$ and allowing for the formation of sharp edges.

As a boundary condition, the edge points of the original straight line were kept fixed, so as to avoid unnecessary strain on the *snake*. The *snakes* resulting from a full run of the algorithm starting from Fig. 2.3.2 are shown in Fig. 2.4.1, layered on top of the original image for ease of visualization.

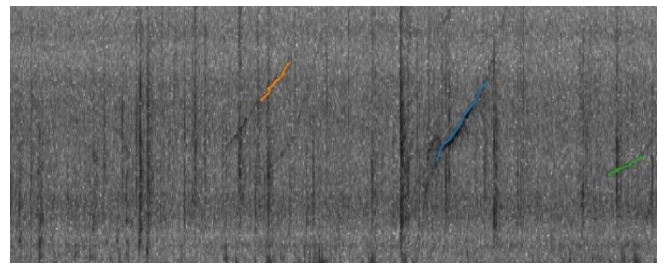

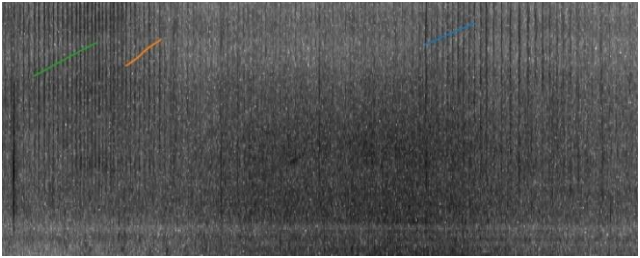
Fig. 2.4.1: *Snakes* overlaying the desired patterns. [color]

Occasionally, a straight line found by the Hough transform algorithm will not cover a real pattern detection, resulting in a meaningless *snake*. Such mistakes can be seen in Fig. 2.4.2, and are the main object of the fourth stage of this study (machine learning classification).

Fig. 2.4.2: A spectrogram image with three false positive events of interest. [color]

## 2.5. Feature vectors

The occurrence of false positives, as highlighted in Fig. 2.4.2, indicates that an algorithm based exclusively on the combination of the Hough transform and active contours is not robust enough to detect estuarine dolphin vocalizations in spectrograms of the landscape as intended. For this reason, a classification algorithm based on a supervised learning approach was added to those components and, as a consequence, it was necessary to extract data from the *snakes* obtained via the previous algorithms and organize them into an appropriate training set.

We now consider a *snake* a collection of points $v(i) = (x(i), y(i))$, where $i$ is a discrete variable ranging from $i = 1$ to $N$, where $N$ is the number of points in the *snake*. A value of $N = 50$ was chosen for this research, maintaining a balance between computational time and level of detail in the *snake*.

Six geometric features were obtained from each *snake*: the two coordinates $(\bar{x}, \bar{y})$ of the centroid; the *snake*'s normalized length, taken as the distance between its endpoints divided by a known threshold; the moment of inertia relative to a horizontal axis containing the *snake's* centroid, and its average and relative masses. Each of these features is thereby described in this section.

Eq. 2.5.1 gives the position of the centroid, where $x(i)$ and $y(i)$ represent the x and y coordinates, in pixels, of the i-th point on the *snake*. It is important to note that the x (horizontal) axis in the spectrogram image corresponds to a sequence of time intervals Δt where the spectrogram images were computed, and has no influence on whether a detection by the Hough algorithm is a true positive or not. It was, however, useful for quick manual identification of the *snakes*. The y (vertical) axis in the spectrogram image should be influential, given that a centroid measurement within this axis describes the average frequency of the corresponding sound.

$$(\bar{x}, \bar{y}) = \frac{1}{N}\sum_{i=1}^{N}(x(i), y(i)) \quad (2.5.1)$$

Length was simply defined as the distance between the endpoints, then normalized by a selected value l equal to the length of the shortest *snake* encountered in a given set of images, and is given in Eq. 2.5.2.

$$L_{snake} = \frac{1}{l}\sqrt{(x(N) - x(1))^2 + (y(N) - y(1))^2} \quad (2.5.2)$$

The next three features make use of the gray level image function I(x, y), defined to return the normalized intensity of the image at the point (x, y), from 0 (black) to 1 (white). Here, an important (albeit counter-intuitive) distinction shows that darker areas in the spectrogram image correspond to higher-intensity frequency patches, but were assigned with low values in the Matplotlib color map. The gray levels of the image were employed to represent the "density" of each point, for purposes of calculating the moment of inertia and a concept of mass, so that the patterns associated with the events of interest correspond to areas of lower image density.

The moment of inertia was calculated with respect to a horizontal axis passing through the snake's centroid, and is therefore defined by Eq. 2.5.3.

$$J_{snake} = \sum_{i=1}^{N} I(x(i), y(i)) \times (y(i) - \bar{y})^2 \quad (2.5.3)$$

With an analogous notion to density as defined by the gray level image function, it becomes possible to calculate the mass of a *snake* with N points, and to further compare it to the density of the spectrogram image as a whole, shaped a × b (pixels), used as a discretization of the x and y axes. The motivation for this comparison comes from the assumption that true detections will be darker than the mean gray level of the image. This line of reasoning will be further detailed in section 3.

From the aforementioned interpretations follow the definitions for the masses of the snake: average $M_{snake}$ and relative $RM_{snake}$, respectively shown in Eqs. 2.5.4 and 2.5.5, where v(i) are points of the snake, as considered earlier in this section, and the sum of I(i, j) refers to adding up the gray level of every pixel in the spectrogram image.

$$M_{snake} = \frac{1}{N}\sum_{i=1}^{N} I(v(i)) \quad (2.5.4)$$

$$RM_{snake} = \frac{M_{snake}}{\frac{1}{ab}\sum_{i=1}^{b}\sum_{j=1}^{a} I(i,j)} \quad (2.5.5)$$

## 2.6. The Random Forest classifier

A random forest classifier (Breiman, 2001) is a collection of decision tree-like classifiers (Breiman et al., 1984) averaged on random subsets of the original data. The original implementation of a single decision tree involves branching decisions such as, for instance, deciding to classify a detection as true if its relative density goes below a certain cutoff.

In decision trees (James et al., 2017), two of the main criteria for evaluating where a split is necessary (e.g. a branching decision at specific values of a discrete feature, or thresholds of a continuous one) are calculating the entropy and the Gini index of a split.

For a categorical feature $Y$ with $K$ values, the entropy of a set $D$ of observations is defined in Eq. 2.6.1.

$$I(D) = -\sum_{k=1}^{K} \frac{|D_k|}{|D|} \log_2 \frac{|D_k|}{|D|} \quad (2.6.1)$$

Here, $D_k$ is the subset of $D$ where $Y = k$ and $|D|$ is the number of elements of the set $D$. From this definition, we can calculate the gain ratio of a split into subsets $D^1, \dots, D^m$ by Eq. 2.6.2.

$$I_{gain} = \frac{I(D) - \sum_{j=1}^{m} \frac{|D^j|}{|D|} I(D^j)}{-\sum_{j=1}^{m} \frac{|D^j|}{|D|} \log_2 \frac{|D^j|}{|D|}} \quad (2.6.2)$$

Therefore, we compute the gain ratio of every possible split in the decision tree, and choose the one with largest ratio.

Alternatively, we can compute the Gini index of a region $D$ through Eq. 2.6.3.

$$g(D) = \sum_{k=1}^{K} \frac{|D_k|}{|D|} \left(1 - \frac{|D_k|}{|D|}\right) \quad (2.6.3)$$

The parameter for evaluating a split into sets $D^1$ and $D^2$ is thereby defined in Eq. 2.6.4.

$$g_{gain} = g(D) - \frac{|D^1|}{|D|} g(D^1) - \frac{|D^2|}{|D|} g(D^2) \quad (2.6.4)$$

A random forest trains several decision trees on various subsets of the dataset and averages their decisions, an effective method to reduce over-fitting. The implementation used in this study is the Python version made by scikit-learn, which is itself based on (Breiman, 2001) and (Geurts et al., 2006).

The definition of a random forest is an ensemble of tree-structured classifiers $h(x, \Theta_k)$, where $\{\Theta_k\}_{k=1,\dots}$ are random vectors that are independent and identically distributed. Then, the final decision made by the random forest is the most popular decision among the classifiers at the input $x$.

A classic implementation of a random forest, such as the one used for the classification of the estuarine dolphin vocalization detections, randomizes the features at each cutoff, resulting in slightly different classifiers every time the algorithm is run. Therefore, the final accuracy value obtained may fluctuate by small amounts.

It can be shown that a random forest always converges, and so the randomization of the features is an effective way to prevent overfitting. In (Breiman, 2001), the features are selected via an "out-of-bag" method.

The random forest was deemed an adequate choice for this study based on the presence of various natural thresholds and cutoff points for the features, as will become clear during the discussion of the results of the proposed method.

# 3. Results

The relevant patterns identified in the available dataset, as well as the optimal accuracy obtained with the previously discussed Random Forest classifier, are the focus of this section.

The target variable for the classifier is binary, indicating whether a *snake* detection truly corresponds to the characteristic pattern of interest. It will henceforth be referred to simply as "target." Examples of correct and incorrect detections are shown in Fig. 3.1. The incorrect *snake* is shown in red, near the upper left corner.

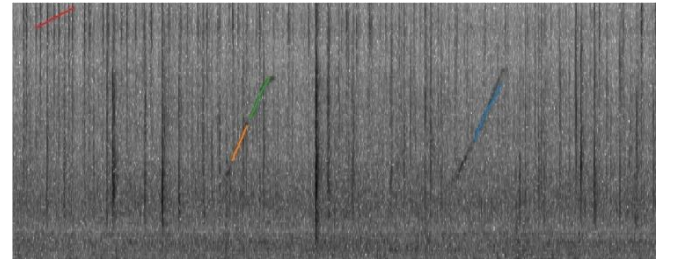

Fig. 3.1: Examples of three correct detections and a false positive one. [color]

## 3.1. Noteworthy patterns in the dataset

Through manual classification of the available *snakes*, three main patterns were identified and represented by unique variables, thus aiding the construction of the classifier. The validity of such patterns was subsequently tested.

The first such pattern is referred to as the relative mass of the *snake*, defined previously in section 2.5 through Eqs. 2.5.4 and 2.5.5. This parameter is a comparison between the gray levels of the points covered by the *snake* and the average gray level of the whole spectrogram image. It is expected that, for a true detection, the *snake* covers darker points, which correspond to higher frequency intensities in the spectrogram, and therefore has a low relative mass. An arbitrary threshold was selected through careful observation of the dataset, separating "low-" from "high-density" snakes in a binary variable. Thus, it was defined that a "low-density" *snake* (that is, a *snake* in a low-density region) satisfies
$$RM_{snake} \leq 0.8.$$

A graphical representation of the number of manually classified detections, seen in Fig. 3.1.1, confirms the expected behavior. The columns on the graph, as labeled, indicate how many true or false detections are "low-" or "high-density" curves.

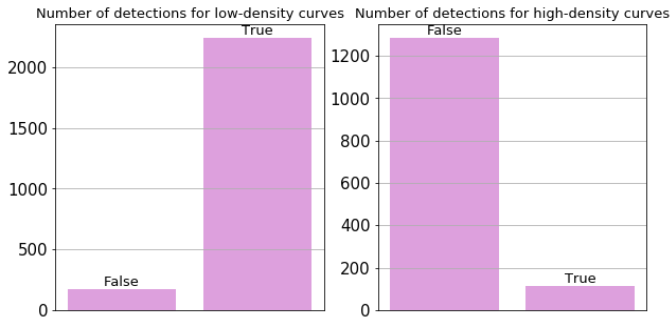

Fig. 3.1.1: Instances of the target valuable in relation to the binary density feature.

A second pattern of interest is the length of the *snake*. It arises from the observation that, since false detections tend to be a random collection of alligned darker points, most of them will not be longer than average. Meanwhile, many true detections may cover a large portion of the image. Examples of this behavior in false detections are shown in Fig. 3.1.2.

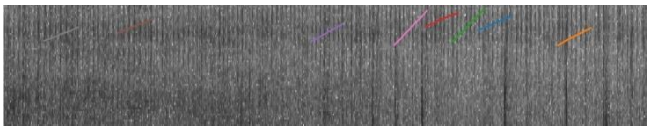

Fig. 3.1.2. Several false detections within a single image. [color]

With length defined by Eq. 2.5.2, a threshold was chosen for a desired length. It is expected that most long curves will be true detections, while short curves may not have any detectable bias. It was established that a "long" curve satisfies

$$L_{snake} \geq 3.$$

A graphical analysis of this new feature confirms the assumed patterns, as shown in Fig. 3.1.3. Most "long" curves are, therefore, true detections, which may serve as a useful cutoff point for the random forest classifier.

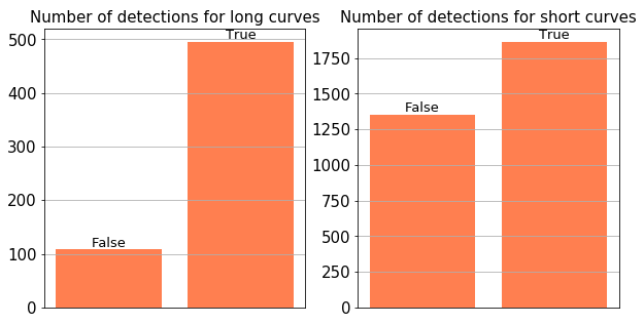

Fig. 3.1.3: Instances of the target valuable in relation to the binary length feature.

A third and final pattern arises from a particular collection of images containing several counter-examples for the density pattern. The presence of a constant low-frequency disturbance in the data, likely caused by passing ships, may generate lower density patches in the image, leading false detections to have a lower relative density despite their usual patterns. *Snakes* that exhibit this pattern are shown in Fig. 3.1.4.

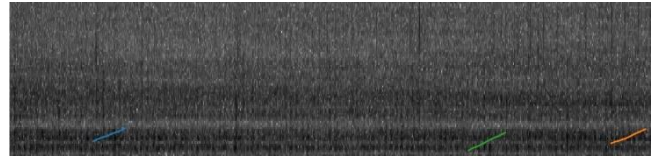

Fig. 3.1.4: Examples of false detections due to a low-density disturbance. [color]

Hence, verifying if the *snake* corresponds to a range of low frequencies in the spectrogram is very important: in case it does, the detected event is likely to be a false positive. Other observations of the training data indicate that very few true detections lie in the lower sections, mostly due to the high-frequency nature of the pattern of interest.

This analysis leads to a third classification of curves as "low-" or "high-frequency". A "low-frequency" curve with N points is defined to be one such that its centroid satisfies

$$\bar{y} \leq 510.$$

"High-frequency" curves are exempt from the considerations above, and thus expected to be free of any bias. This pattern is then visualized in Fig. 3.1.5.

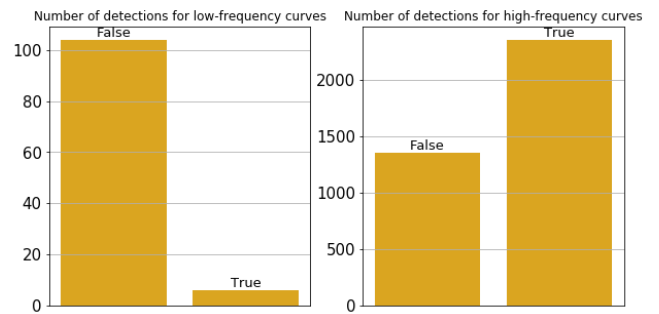

Fig. 3.1.5: Instances of the target valuable in relation to the binary frequency feature.

A full correlation matrix for all defined features, including both the new patterns and the ones previously discussed in section 2.5, is shown in Fig. 3.1.5. It also confirms the previously predicted biases in the newly defined binary patterns.

Correlation matrix for the training dataset

|  | avg_density | avg_x | avg_y | inertia | length | relative_density | low_density | long | low | target |
|---|---|---|---|---|---|---|---|---|---|---|
| avg_density | 1.0 | 0.01 | -0.59 | -0.12 | -0.23 | 0.93 | -0.79 | -0.19 | -0.04 | -0.7 |
| avg_x | 0.01 | 1.0 | 0.0 | -0.03 | -0.03 | -0.01 | 0.01 | -0.02 | 0.02 | 0.01 |
| avg_y | -0.59 | 0.0 | 1.0 | 0.05 | 0.08 | -0.65 | 0.65 | 0.07 | 0.44 | 0.45 |
| inertia | -0.12 | -0.03 | 0.05 | 1.0 | 0.76 | -0.19 | 0.23 | 0.6 | -0.1 | 0.29 |
| length | -0.23 | -0.03 | 0.08 | 0.76 | 1.0 | -0.26 | 0.21 | 0.77 | -0.12 | 0.24 |
| relative_density | 0.93 | -0.01 | -0.65 | -0.19 | -0.26 | 1.0 | -0.87 | -0.21 | -0.04 | -0.8 |
| low_density | -0.79 | 0.01 | 0.65 | 0.23 | 0.21 | -0.87 | 1.0 | 0.17 | 0.09 | 0.84 |
| long | -0.19 | -0.02 | 0.07 | 0.6 | 0.77 | -0.21 | 0.17 | 1.0 | -0.07 | 0.18 |
| low | -0.04 | 0.02 | 0.44 | -0.1 | -0.12 | -0.04 | 0.09 | -0.07 | 1.0 | -0.2 |
| target | -0.7 | 0.01 | 0.45 | 0.29 | 0.24 | -0.8 | 0.84 | 0.18 | -0.2 | 1.0 |

Fig. 3.1.5: Correlation matrix for the training dataset. [color]

The last row of the correlation matrix shows the correlations of the calculated features with the target variable. The patterns discussed above are thus confirmed: The binary feature "low-density" exhibits a very high correlation (0.84) with the target variable. The "low-frequency" feature, denoted simply as "low" in Fig. 3.1.5, has a negative correlation with the target variable, as expected.

Many of the features are cross-correlated. There are expected correlations between the binary features and their respective continuous ones. We also observe other correlations, e.g. between length and moment of inertia.

## 3.2. Performance of the Random Forest classifier

In possession of three new binary features associated to the observed patterns, a Random Forest classifier was initialized and subsequently optimized by a grid search method, following the scikit-learn implementation for Python, based on (Bergstra and Bengio, 2012).

A grid search consists of a brute force test of many combinations of the classifier's own hyperparameters within the training dataset. The hyperparameters chosen to be varied in the grid search process were two: the number of estimators within the Random Forest algorithm and the criterion (Gini or entropy) used in the creation of new branches.

Two different implementations of the Random Forest algorithm were attempted. The first made use of all available features, which can be seen in the correlation matrix of Fig. 3.1.6, except for $avg\_x$, as previously discussed. The second one removed a continuous feature whenever a binary pattern had been defined from it. Thus, the three features defined in section 3.1 were kept, and their corresponding features "avg_density," "length" and "avg_y" were removed.

Applications of those two classifiers after their optimization via the grid search method yielded predictions for the test dataset. The first implementation, with all the features, resulted in a total accuracy of 0.977. In the second implementation, the achieved accuracy was 0.966. This discrepancy held through several experimental re-runs of the algorithm, with the accuracies fluctuating between these values by negligible deviations, not large enough to cover the difference between them. The second implementation was thereby discarded.

The confusion matrix for the first implementation is shown in the following table.

Table 3.2.1: Confusion matrix for the classifier.

|  |  | Predicted label | |
|---|---|---|---|
|  |  | 0 | 1 |
| True label | 0 | 367 | 13 |
|  | 1 | 9 | 566 |

The confusion matrix shows a very small number of false positives and false negatives. Their rate was similar, with 9 false negatives and 13 false positives, but very small compared to the number of correct detections (933 in total) by the random forest classifier.

## 3.3. Computational performance of the algorithm

The algorithm as discussed in the previous sections was run on 110 different .wav files of the original dataset, consisting of 6600 images and totaling 5 hours and 30 minutes of audio.

Generation of the Frangi maps took 10 minutes in total. The Hough transform algorithm for straight line detection was faster, averaging at 1 minute for all images. The *snake* algorithm took most of the runtime, averaging 15 minutes per image. The full implementation and testing of both Random Forest classifiers took, on average, 23 minutes per full run, after the training dataset had already been built.

All computations were run on a laptop operating Windows 10, with 8GB RAM available and an Intel(R) Core(TM) i7-5500U CPU @ 2.40 GHz processor, on the native Python 3.6 IDLE environment.

## 4. Conclusions

In this paper, a method for detecting estuarine dolphin vocalizations without intervention of a human operator was developed.

The proposed method is based on the application of a probabilistic Hough transform over the Frangi ridge map of a spectrogram image. Furthermore, the line segments obtained are run through an active contour algorithm and geometric features are extracted. The resulting curves and their features form a dataset for the application of the random forest algorithm.

All data used in our research consists of feature vectors extracted from curves identified in spectrogram images from the underwater soundscape where the referred species is found.

The proposed algorithm offers the following advantages in relation to current methods in the literature: it requires no human interference aside from data collection and organization; it automatically classifies identified patterns as estuarine dolphin vocalizations or false detections; it is based only on the approximate linearity of the pattern, and therefore still performs well in spectrogram images containing high noise levels.

The best implementation of the random forest classifier obtained a raw accuracy of 0.977, with a false positive rate of 0.034 and a false negative rate of 0.016, as taken from Fig. 3.2.1.

For future research, we intend to investigate the performance of the algorithm in spectrogram images that contain not only estuarine dolphin vocalizations, but include other characteristic acoustic events, both anthropogenic and natural.

We intend to test our algorithm in images with lower signal-to-noise ratio and extend the method implemented in this article to be able to recognize vocalizations from other species of cetaceans.

**Declaration of Interest**

The authors declare that they have no competing interests.

**Acknowledgements**

This work was partly supported by CNPq grant 303992 / 2017-4 and FAPESP grants CEPID-Shell-RCGI 2014/50279-4.